\documentclass[%
 reprint,
 amsmath,amssymb,
 aps,
]{revtex4-1}

\usepackage{graphicx}
\usepackage{dcolumn}
\usepackage{bm}

\begin{document}

\preprint{APS/123-QED}

\title{Consistent forcing scheme in the cascaded lattice Boltzmann method}

\author{ Linlin Fei$^{1}$, K. H. Luo$^{1,2}$\footnote{Corresponding author: K.Luo@ucl.ac.uk}}
\affiliation{$^1$ Center for Combustion Energy; Key laboratory for Thermal Science and Power Engineering of Ministry of Education, Department of Thermal Engineering, Tsinghua University, Beijing 100084, China \\
	$^2$ Department of Mechanical Engineering, University College London, Torrington Place, London WC1E 7JE, UK\\
}

\date{\today}

\begin{abstract}
In this paper, we give a more pellucid derivation for the cascaded lattice Boltzmann method (CLBM) based on a general multiple-relaxation-time (MRT) frame through defining a shift matrix. When the shift matrix is a unit matrix, the CLBM degrades into an MRT LBM. Based on this, a consistent forcing scheme is developed for the CLBM. The applicability of the non-slip rule, the second-order convergence rate in space and the property of isotropy for the consistent forcing scheme is demonstrated through the simulation of several canonical problems. Several other existing force schemes previously used in the CLBM are also examined. The study clarifies the relation between MRT LBM and CLBM under a general framework.
\begin{description}
	\item[PACS numbers]
	47.11.-j, 05.20.Dd
\end{description}
\end{abstract}                             

\maketitle


\section{Introduction}

The lattice Boltzmann method (LBM), based on the simplified kinetic models, has gained remarkable success for the simulation of complex fluid flows and beyond,
with applications (but not limited to) to micro flows, flows in porous media, turbulence, and multiphase flows \cite{qian1995recent,chen1998lattice,succi2001lattice,shan1993lattice,guo2013lattice,Gong2016Wetting}. The LBM solves a specific discrete Boltzmann equation for the distribution functions, designed to recover the Navier-Stokes (N-S) equations in the macroscopic limit. The meso-scale nature of LBM allows its natural incorporation of micro and meso-scale physics, while the highly efficient algorithm makes it afforadable computationally \cite{li2016lattice}.

In the standard ``collision-streaming" LBM algorithm, the simplest collision operator is the Bhatnagar-Gross-Krook (BGK) or single-relaxation-time (SRT) operator, which relaxes all the distribution functions to their
local equilibrium counterparts at a common rate and the relaxation rate is related to the kinematic viscosity \cite{qian1992lattice}. The multiple-relaxation-time (MRT) operator is another extensively used operater \cite{d1994generalized}, in which the collision is executed in the moment space and the relaxation rates for different moments can be different. More recently, a central-moments-based or cascaded operator was proposed by Geier \emph{et al}. \cite{geier2006cascaded}. In the cascaded lattice Boltzmann method (CLBM), the collision is carried out in the space of central moment rather than that of raw moment as in the MRT LBM.  Compared with the BGK operator, the MRT and cascaded operators can increase the numerical stability significantly \cite{geier2006cascaded,lallemand2000theory,lycett2016cascaded,fei2016thermal}. Athough the collision steps in these LBMs are quite different, the streaming steps are carried out in the same way by streaming the post-collision distributions to their neighbors. It should be noted that other collision operaters, like the two-relaxation-time (TRT) operator \cite{ginzburg2005equilibrium,ginzburg2008two} and the entropic operator \cite{ansumali2003minimal,ansumali2000stabilization} are also very popular in the lattice Boltzmann community.

In many fluid systems, an external or internal force field plays an important role in the flow behaviours. To incorporate the force effect, different force treatments have been proposed in the literature \cite{shan1993lattice,he1999lattice,buick2000gravity,ladd2001lattice}. In 2002, Guo \emph{et al} analyzed the discrete lattice effects on the forcing scheme and 
developed a representation of the forcing term \cite{guo2002discrete}. Guo \emph{et al}. then extended the method to the MRT LBM in 2008 \cite{Guo2008Analysis}. Up to now, the method by Guo \emph{et al} has been widely used in the LBM simulation. For the CLBM simulation, there is still no commonly used forcing scheme, while one scheme by method of central moments has been  proposed by Premnath \emph{et al} .\cite{premnath2009incorporating}. In Ref. \cite{lycett2014multiphase},  Lycett-Brown and Luo incorporate the forcing scheme for the BGK LBM into the CLBM directly. As analysed by De Rosis \cite{de2017alternative}, the method proposed by Premnath \emph{et al} may encounter cumbersome practical implementations. Based on the central moments of a discrete equilibrium, a forcing scheme has been developed in \cite{de2017alternative}. 

However, there is still no analysis about whether these forcing schemes in the CLBM are consistent with the forcing scheme
in the MRT LBM \cite{Guo2008Analysis} and the original scheme proposed by Guo \emph{et al} in the BGK LBM \cite{guo2002discrete}.
In this paper, we propose a more pellucid derivation for the CLBM by defining a shift matrix. This definition clarifies
the relationship between MRT LBM and CLBM. Based on this frame, we present a consistent forcing scheme in CLBM, and show that the privious methods in Refs. \cite{premnath2009incorporating,lycett2014multiphase,de2017alternative} are not consistent. The rest of the paper is structured as follows: Section \ref{sec.2} gives the new derivation for the cascaded LBM and presents the consistent forcing scheme. Section \ref{sec.3} presents a short analysis for the privious forcing schemes. Numerical verifications are presented in Sec. \ref{sec.4}. Finally, conclusions of this work are made in Sec. \ref{sec.5}.
\section{CASCADED LBM AND CONSISTENT FORCING SCHEME}\label{sec.2}
\subsection{Cascaded LBM}\label{sec.2a}
Without losing the generality, the D2Q9 lattice \cite{qian1992lattice} is adopted here. The lattice speed  $ c=\Delta{x}/\Delta {t}=1 $ and the lattice sound speed $ c_{s}=1/\sqrt{3} $ are adopted, in which $ \Delta{x} $ and $ \Delta {t} $ are the lattice spacing and time step. The discrete velocities $ {{\bf{e}}_i} = \left[ {\left| {{e_{ix}}} \right\rangle ,\left\langle {{e_{iy}}} \right|} \right] $ are defined as 
\begin{equation}\label{e1}
\begin{array}{l}
\left| {{e_{ix}}} \right\rangle  = {[0,1,0, - 1,0,1, - 1, - 1,1]^\top}, \\ 
\left| {{e_{iy}}} \right\rangle  = {[0,0,1,0, - 1,1,1, - 1, - 1]^\top}. \\ 
\end{array}
\end{equation}
where $ i = 0...8 $, ${\left|  \cdot  \right\rangle }$ denotes a nine-dimensional colunm vector, and the superscript $ \top $ denotes the transposition. 

Here we propose a new derivation for the CLBM, which is different from and more intelligible than that given by Geier \emph{et al} \cite{geier2006cascaded}. We first define the raw and central velocity moments of the discrete distribution function (DF) ${{f_i}}$,
\begin{equation}\label{e2}
\begin{array}{l}
{k_{{mn}}} = \left\langle {{f_i}\left| {e_{ix}^me_{iy}^n} \right.} \right\rangle , \\ 
{{\tilde k}_{{mn}}} = \left\langle {{f_i}\left| {{{({e_{ix}} - {u_x})}^m}{{({e_{iy}} - {u_y})}^n}} \right.} \right\rangle , \\ 
\end{array}
\end{equation}
where $ {u_x} $ and $ {u_y} $ are the horizontal and vertical velocity components. The equilibrium values $ k_{_{{mn}}}^{eq} $ and $ \tilde k_{{mn}}^{eq}$ are defined
analogously by using the discrete equilibrium distribution function (EDF) $ {f_i^{eq}} $ in Eq. (\ref{e2}). In the previous CLBM, the recombined raw moments are adopted,
\begin{equation}\label{e3}
\left| {{T_i}} \right\rangle  = 
\left[ {{k_{00}},{k_{10}},{k_{01}},{k_{20}} + {k_{02}},	{k_{20}} - {k_{02}},{k_{11}},{k_{21}},{k_{12}},{k_{22}}} \right]^\top,
\end{equation}
so do the recombined central moments $ \tilde T_i$. The transformantion from the discrete DF to its raw moments can be realized through a transformation matrix $ {\bf{M}} $, and the shift from the raw moments to central moments can be realized though a shfit matrix $ {\bf{N}} $,
\begin{equation}\label{e4}
\begin{array}{l}
\left| {{T_i}} \right\rangle  = {\bf{M}}\left| {{f_i}} \right\rangle , \\ 
\left| {{{\tilde T}_i}} \right\rangle  = {\bf{N}}\left| {{T_i}} \right\rangle . \\ 
\end{array}
\end{equation}
The explicit forms for $ {\bf{M}}
$ and $ {\bf{N}}
$ can be obtained through the defination in Eqs. (\ref{e2},\ref{e3},\ref{e4}). Explicitly, the transformation matrix 
$ {\bf{M}} $ is expressed as \cite{lycett2014multiphase}
\begin{equation}\label{e5}
{\bf{M}} = \left[ 
\begin{array}{c c c c c c c c c}
1 &1 &1&1&1&1&1&1&1\\
0&1&0&-1&0&1&-1&-1&1\\
0&0&1&0&-1&1&1&-1&-1\\
0&1&1&1&1&2&2&2&2\\
0&1&-1&1&-1&0&0&0&0\\
0&0&0&0&0&1&-1&1&-1\\
0&0&0&0&0&1&1&-1&-1\\
0&0&0&0&0&1&-1&-1&1\\
0&0&0&0&0&1&1&1&1\\
\end{array} 
\right],
\end{equation}
and the shift matrix $ {\bf{N}} $ is given by
\begin{widetext}
	\begin{equation}\label{e6}
	{\bf{N}} = \left[ 
	\begin{array}{c c c c c c c c c}
	1 &0 &0&0&0&0&0&0&0\\
	-{u_x}&1&0&0&0&0&0&0&0\\
	-{u_y}&0&1&0&0&0&0&0&0\\
	u_x^2+u_y^2&-2{u_x}&-2{u_y}&1&0&0&0&0&0\\
	u_x^2-u_y^2&-2{u_x}&2{u_y}&0&1&0&0&0&0\\
	{u_x}{u_y}&-u_y&-u_x&0&0&1&0&0&0\\
	-u_x^2{u_y}&2{u_x}{u_y}&u_x^2&  - {u_y}/2
	
	&- {u_y}/2&-2u_x&1&0&0\\
		-u_y^2{u_x}&{u_y}^2&2{u_x}{u_y}&  - {u_x}/2
	
	& {u_x}/2&-2u_y&0&1&0\\
	u_x^2u_y^2&-2{u_x}u_y^2&-2{u_y}u_x^2&u_x^2/2+u_y^2/2&u_y^2/2-u_x^2/2&4{u_x}{u_y}&- 2{u_y}&- 2{u_x}&1\\
	\end{array} 
	\right].
	\end{equation}
\end{widetext}
In the collision step for the cascaded LBM, the central moments ($ \tilde T_i$) of the discrete DF ${{f_i}}$ are
relaxed to their equilibrium values $ \tilde T_i^{eq} $. Thus the post-collision central moments are
\begin{equation}\label{e7}
\begin{aligned}
\left| {\tilde T_i^*} \right\rangle & = ({\bf{I - S}})\left| {{{\tilde T}_i}} \right\rangle  + {\bf{S}}\left| {\tilde T_i^{eq}} \right\rangle  \\
&= ({\bf{I - S}}){\bf{NM}}\left| {{f_i}} \right\rangle  + {\bf{SNM}}\left| {f_i^{eq}} \right\rangle.  
\end{aligned}
\end{equation}
where $ {\bf{S}} = diag({s_0},{s_1},{s_1},{s_b},{s_2},{s_2},{s_3},{s_3},{s_4})
 $ is a diagonal relaxation matrix. The kinematic and bulk viscosities are related to the relaxation parameters $ {s_2} = 1/(3\upsilon  + 0.5) $ and $ {s_b} = 1/(3\xi  + 0.5) $, respectively.
As recommened in Refs. \cite{geier2006cascaded,premnath2009incorporating,lycett2014multiphase}, the equalibrium central moments of the 
discrete (EDF) $ {f_i^{eq}} $ are set equal to the continuous central moments of the Maxwellian-Boltzmann distribution in continuous velocity space. To be specific,
\begin{equation}\label{e8}
\left| {\tilde T_i^{eq}} \right\rangle  = \left[ {\rho ,0,0,2\rho c_s^2,0,0,0,0,\rho c_s^4} \right]^\top,
\end{equation}
where $ \rho $ is the fluid density, thus the matrix manipulation is not needed for $ \left| {\tilde T_i^{eq}} \right\rangle $. The corresponding discrete EDF $ {f_i^{eq}} $ is in fact a generalized
local equilibrium \cite{Asinari2008Generalized,premnath2009incorporating}. Due to the definitions of the transformation and shift matrices, both of them are reversible (explicit expressions for $ {\bf{M}}^{ - 1} $ and $ {\bf{N}}^{ - 1} $ are given in the Appendix).
The post-collision discrete DF is given by 
\begin{equation}\label{e9}
\left| {f_i^*} \right\rangle  = {{\bf{M}}^{ - 1}}{{\bf{N}}^{ - 1}}\left| {\tilde T_i^*} \right\rangle. 
\end{equation}
In the  streaming step, the post-collision discrete DF in space $ \bf{x} $ and time $ t $ streams to its neighbor in the next time step  as usual
\cite{guo2002discrete,Guo2008Analysis,liu2016non},
\begin{equation}\label{e10}
{f_i}(\textbf{x} + {\textbf{e}_i}\Delta t,t + \Delta t) = f_i^*(\textbf{x},t).
\end{equation}
Using the Chapman-Enskog analysis, the incompressible N-S equaltions can be reproduced in the low-Mach number limit \cite{premnath2009incorporating,lycett2014multiphase}. The 
hydrodynamics variables are obtained as,
\begin{equation}\label{e11}
\rho  = \sum\nolimits_i {{f_i}} ,~~~\rho \textbf{u} = \sum\nolimits_i {{f_i}} {\textbf{e}_i}.
\end{equation}

It can be found that when the shift matrix  $ {\bf{N}} $ is a unit matrix, the CLBM degrades into a non-orthogonal MRT LBM \cite{liu2016non}. Thus the present derivation is based on a general multiple-relaxation-time frame and clarifies the relationship between
the MRT LBM and CLBM.
\subsection{Consistent forcing scheme}\label{sec.2b}
Inspired by the method proposed by Guo \emph{et al} \cite{guo2002discrete,Guo2008Analysis}, to incoporate an external or internal force field $ {\bf{F}} = [{F_x},{F_y}]
 $ into the CLBM , the collision step for central moments in Eq. (\ref{e6}) is modified by
\begin{equation}\label{e12}
\begin{aligned}
\left| {\tilde T_i^*} \right\rangle  &= ({\bf{I - S}})\left| {{{\tilde T}_i}} \right\rangle  + {\bf{S}}\left| {\tilde T_i^{eq}} \right\rangle  + ({\bf{I - S}}/2)\left| {{C_i}} \right\rangle  \\ 
&= {\bf{NM}}\left| {{f_i}} \right\rangle  + {\bf{SNM}}\left| {f_i^{eq}} \right\rangle  + ({\bf{I - S}}/2){\bf{NM}}\left| {{R_i}} \right\rangle   
\end{aligned}
\end{equation}
where $ {{R_i}} $ is the force effect term, which can be obtained 
by \cite{he1998novel},
\begin{equation}\label{e13} 
{{R_i}}  = \frac{{\bf{F}}}{\rho } \cdot \frac{{\left( {{\textbf{e}_i} - \textbf{u}} \right)}}{{c_s^2}}f_i^{eq}.
\end{equation}
When using the generalized local equalibrium, the central moments for $ \left| {{R_i}} \right\rangle $ can be computed as,
\begin{equation}\label{e14}
\left| {{C_i}} \right\rangle  =  [0,{F_x},{F_y},0,0,0,c_s^2{F_x},c_s^2{F_y},0]^\top,
\end{equation}
so the explicit formulation and matrix manipulation for $ \left| {{R_i}} \right\rangle $ in Eq. (\ref{e12}) are not needed in the practical implement. Then the fluid velocity is defined by,
\begin{equation}\label{e15}
\rho {\bf{u}} = \sum\nolimits_i {{f_i}} {{\bf{e}}_i} + \frac{{\Delta t}}{2}{\bf{F}}.
\end{equation}

\textit{Remark 1}. In Eq. (\ref{e12}), the forcing effects in the present scheme are considered by means of central moments, which is compatible with the basic ideology (collision in the 
central moments space) in the CLBM.

\textit{Remark 2}. When the shift matrix  $ {\bf{N}} $ in Eq. (\ref{e12})
is a unit matrix, the CLBM with the present forcing scheme will degrade
into the MRT LBM proposed by Liu \emph{et al}. \cite{liu2016non} with some
high-order terms. It is known that the method of Liu \emph{et al}. is equivalent to the method of Guo \emph{et al}. in \cite{Guo2008Analysis}. 

\textit{Remark 3}. In the orginal forcing scheme proposed by Guo \emph{et al}. \cite{guo2002discrete}, the forcing effect term is defined by $ {R_{Gi}}
 = {w_i}\left[ {\left( {{\textbf{e}_i} - \textbf{u}} \right)/c_s^2 + \left( {{\textbf{e}_i} \cdot \textbf{u}} \right){\textbf{e}_i}/c_s^4} \right]{\bf{F}} $. It is easy to find that the forcing effect term in Eq. (\ref{e13}) is equivalent to $ {R_{Gi}} $ plus some high-order terms. The constraint conditions for the forcing effect term (see Eq. (7) in \cite{guo2002discrete}) are also satisfied in the present scheme. In particular, when all the parameters in the matrix  $ {\bf{S}} $ are set equal to $ {s_2}
  $, the CLBM with the present forcing scheme will degrade into the BGK LBM with forcing scheme by  Guo \emph{et al}. in \cite{guo2002discrete} with a generalized local equilibrium.
  
\textit{Remark 4}. It is also found (see in Sec. \ref{sec.4}) that the zero-slip velocity boundary condition
for the half-way bounce-back rule ($ {s_3} = (16 - 8{s_2})/(8 - {s_2})
 $) discussed in \cite{Guo2008Analysis,ginzburg2003multireflection} is also  applicable to the present forcing scheme.
  
From the above, we name the present forcing scheme as a consistent scheme in the CLBM.

\section{OTHER FORCING METHODS}\label{sec.3}
In this section, several other methods to incorporate forcing effects into the CLBM in the literature are summerized. To show the
inconsistencies in these methods, they are all written in the general multiple-relaxation-time frame proposed
in Sec. \ref{sec.2a}.
\subsection{Forcing scheme by Premnath \emph{et al}.}\label{sec.3a}
In 2009, Premnath \emph{et al}. \cite{premnath2009incorporating} proposed a forcing scheme to incorporate forcing terms into CLBM. Inspired by He \emph{et al}. \cite{he1998novel}, they proposed a change of continuous distribution function $ f
 $ due to the presence of a force field,
\begin{equation}\label{e16}
\Delta f = \frac{{\bf{F}}}{\rho } \cdot \frac{{\left( {{e_i} - u} \right)}}{{c_s^2}}{f^M},
\end{equation}
where $ {f^M} $ is the Maxwellian-Boltzmann distribution in continuous velocity space. The central moments of  $ \Delta f $ is then incorporated into the collision stage in central moments by, 
\begin{equation}\label{e17}
\left| {\tilde T_i^*} \right\rangle {\rm{ }} = ({\bf{I}} - {\bf{S}})\left| {{{\tilde T}_i}} \right\rangle  + {\bf{S}}\left| {\tilde T_i^{eq} - {C_{Pi}}/2} \right\rangle. 
\end{equation}
The discrete counterpart of $ \Delta f $ is also needed to obtain the post-collision discrete DF,
\begin{equation}\label{e18}
\left| {f_i^*} \right\rangle  = {{\bf{M}}^{ - 1}}{{\bf{N}}^{ - 1}}\left| {\tilde T_i^*} \right\rangle  + \left| {\Delta {f_i}} \right\rangle  
\end{equation}
The fluid velocity is defined as in Eq. (\ref{e15}). 

It should be noted that the original derivation in \cite{premnath2009incorporating} is tedious, and $ {C_{Pi}} $ and $ {\Delta {f_i}} $ are corresponding to $ {{\hat \sigma }_{{x^m}{y^n}}}
 $ and $ {S_a} $ in \cite{premnath2009incorporating}, respectively. Though the method
 is compatible with the central-moment-based collision operator, the explicit formulations of $ {\Delta {f_i}} $ and its 
 raw moments are needed, which makes the practical implementations cumbersome \cite{de2017alternative}. Moreover, in their definition for the central moments of $ \Delta f $, they removed the high-order nonzero terms arbitraily,
\begin{equation}\label{e19}
\left| {{C_{Pi}}} \right\rangle  = [0,{F_x},{F_y},0,0,0,0,0,0]^\top.
\end{equation}

Though they think that the high-order terms do not affect consistency, we certainly see some inconsistencies. For example, the first element in $ {\Delta {f_i}} $ and $  {R_{Gi}} $ are apparently inconsistent,
\begin{equation}\label{e20}
\begin{array}{l}
\Delta {f_0} =  - 2{F_x}{u_x} - 2{F_y}{u_y} + O({u^3}),  \\ 
{R_{G0}} =  - 3{F_x}{u_x} - 3{F_y}{u_y},\\ 
\end{array}
\end{equation}
which will affect numerical performaces (see in  Sec. \ref{sec.4}).
\subsection{Forcing scheme by Lycett-Brown and Luo}\label{sec.3b}
Cascaded LBM was first used to simulated multiphase flows by Lycett-Brown \emph{et al}. \cite{lycett2014multiphase} in 2014. In their method, three forcing schemes, the Shan-Chen method \cite{shan1993lattice}, the EDM method \cite{kupershtokh2009equations} and Guo method \cite{guo2002discrete} were adopted directly in the CLBM.

 As discussed in the literature \cite{huang2011forcing,li2012forcing}, both the Shan-Chen method and EDM method obtain some additional terms in the recovered macroscopic equations. These addtitional terms may have some positive effets on the numerial performance of the Shan-chen model \cite{shan1993lattice}, but it is not sensible to use the Shan-Chen method and EDM method in the CLBM directly for general flows. In the present work, we only consider the CLBM with the forcing scheme of Guo \emph{et al}. Thus the collision stage in central moments can be written as,
 \begin{equation}\label{e21}
\left| {\tilde T_i^*} \right\rangle  = ({\bf{I}} - {\bf{S}})\left| {{{\tilde T}_i}} \right\rangle  + {\bf{S}}\left| {\tilde T_i^{eq}} \right\rangle  + (1 - {s_2}/2){\bf{NM}}\left| {{R_{Gi}}} \right\rangle, 
 \end{equation}
while the fluid velocity is also defined as in Eq. (\ref{e15}).
\subsection{Forcing scheme by De Rosis}\label{sec.3c}
Recently, De Rosis proposed an alternative method to incorporate forcing effects into the CLBM. The collision stage in central moments is,
 \begin{equation}\label{e22}
\left| {\tilde T_i^*} \right\rangle  = ({\bf{I}} - {\bf{S}})\left| {{{\tilde T}_i}} \right\rangle  + {\bf{S}}\left| {\tilde T_i^{eq}} \right\rangle  + \frac{1}{2}\left| {{\xi _i}} \right\rangle,
\end{equation}
where, $ {\xi _i} $ is the central moment of the forcing effect term, and the fluid velocity is also defined as in Eq. (\ref{e15}).

Unfortunately, there is a typographical error in Eq. (16) of the paper \cite{de2017alternative}. Particularly, the sign in front of $ {\xi_i}/2 $ is not correct. In this method, the forcing term is defined by the
truncated local equilibrium DF, which gives a lot of velocity terms in $ {\xi _i} $ (see Eq. (15) in \cite{de2017alternative}). Due to the definition of central moments, it is not recommended to include velocity terms in $ {\xi _i} $. Thus there are some spurious effects in this method. It is also noted that the computational load for $ {\xi _i} $ is much higher than that of $ {C _i} $ in Eq. (\ref{e14}). Comparing Eq. (\ref{e22}) with Eq. (\ref{e12}), it is seen that the relaxation rate for each element of $ {\xi _i} $ is 1.0 in this method, which is not consistent with the
multiple-ralaxation-time ideology in the CLBM.

\textit{Remark 5}. In the CLBM, the first three central moments are conserved moments, corresponding to conservations of mass and  momentum. Thus  
 the first two parameters ($ {s_0} $ and $ {s_1}
 $) in the relaxation matrix $ {\bf{S}} $ can be chosen freely. This property 
 is retained in the present forcing scheme and the method by Premnath \emph{et al}., because the relaxation 
 matrix acts on the forcing effect terms in these two methods (see Eq. (\ref{e12}) and Eq. (\ref{e17})). However, $ {s_1} $ need to be set equal to $ {s_2} $ in method by Lycett-Brown and Luo, and to be 1.0 method by De Rosis, to guarantee the  conservation of momentum.

\textit{Remark 6}. If $ {s_3} $ is chosen to be 2.0, the forcing effect on the third-order central moment in Eq. (\ref{e11}) is removed. Thus the present scheme
degrades into the forcing scheme by Premnath \emph{et al}. only when $ {s_3}=2.0 $. Similarly, only when $ {s_3}=1.0 $, the difference between the forcing scheme by De Rosis and the present scheme can be removed. And in the BGK limitation, all the parameters are equal to $ {s_2} $, at which the present scheme degrades into the method by Lycett-Brown and Luo.

\textit{Remark 7}. In 2015, an improved forcing scheme for the pseudopotential model in multiphase flow was proposed by Lycett Brown and Luo \cite{Lycett2015Improved}. The imporved forcing scheme was then incorporated into the CLBM for multiphase flow with large-density-ratio at high  Reynolds and Weber numbers \cite{lycett2016cascaded}. The basic philosophy of the improved forcing scheme is making artificial errors in the pressure tensor to counteract the lack of thermodynamical consistency in the original pseudopotential model. Thus it is not suitable for general flows with a force field. Besides, another simple forcing term was used in CLBM to simulate turbulent channel flow in 2011 \cite{freitas2011analysis}. However, as analysed by Guo \emph{et al}. \cite{guo2002discrete}, the method used in \cite{freitas2011analysis} can not recover the accurate macroscopic equations with a spatial and temporal variational force field in the BGK LBM, not to mention in the CLBM.

\section{NUMERICAL SIMULATIONS}\label{sec.4}
In these section, we conduct several benchmark cases to verify the consistent forcing scheme. The other three methods mentioned in Sec. \ref{sec.3} are also used to validate our arguments. The three methods and present method are denoted by $ {M_1} $, $ {M_2} $, $ {M_3} $ and $ {M_p} $, respectively. In the simulation, $ {s_1} $ is set to be  $ {s_2} $ in $ {M_2} $, but to be 1.0 in other methods.

\subsection{Steady Poiseuille flow}\label{sec.4a}
The first problem considered is a steady Poiseuille flow driven by a constant body force $ \textbf{F} $. The flow direction is set to be 
positive direction of the $ x $ axial, thus $ {\bf{F}} = [{F_x},0] $. The
analytical solution for a channel of width $ 2L $ is,

 \begin{equation}\label{e23}
{{\bf{u}}_a} = \left[ {\frac{{{F_x}}}{{2\nu }}(1 - \frac{{{y^2}}}{{{L^2}}}),~0} \right].
\end{equation}

The periodic boundary conditions are used in the flow direction, while the standard half-way bounce-back boundary scheme is used for non-slip boundary conditions at the walls. Due to the simple flow property, the length of the channel is set to $ 3\Delta x $ to save the computational load. 

\begin{figure}
	\includegraphics[width=0.4\textwidth]{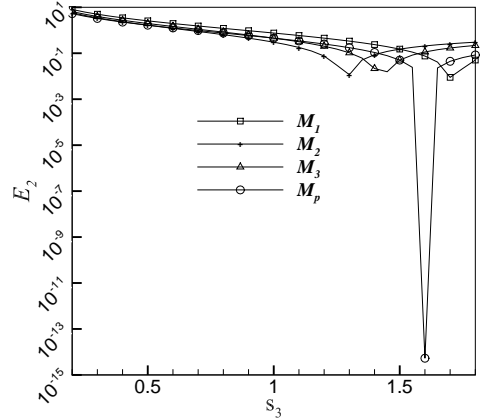}
	\caption{\label{FIG1} $ {E_2} $ changes with   $ {s_3} $ for different methods. }
\end{figure}

As analysed by previous researchers \cite{ginzburg2003multireflection,Guo2008Analysis}, when the relaxation rate for the energy flux is chosen to be $ {s_3} = (16 - 8{s_2})/(8 - {s_2}) $, no numerical slips occurs in the Poiseuille flows for the MRT LBM. To check its applicability in the CLBM with the present forcing scheme, we first choose kinematic viscity $ \upsilon  = 0.5 $, $ {F_x} = 0.01 $, and
only three nodes are used to cover the channel width ($ 2L = 3\Delta x
 $) . We change $ {s_3} $ from 0.2 to 1.8 with a 0.05 interval, and the other parameters are set equal to $ {s_2} $. The residual error $ {E_R} < {1\times10^{ - 9}} $ is used as the convergent criterion, and the relative error  $ {E_2} $ is calculated for the following analysis,
 \begin{equation}\label{e24}
 {E_R} = \sqrt {\frac{{\sum {{{({\mathbf{u}_{(t + 1000\delta t)}} - {\mathbf{u}_t})}^2}} }}{{\sum {\mathbf{u}_{(t + 1000\delta t)}^2} }}},~~ {E_2} = \sqrt {\frac{{\sum {{{(\mathbf{u} - {\mathbf{u}_a})}^2}} }}{{\sum {{\mathbf{u}_a}^2} }}}.
 \end{equation}
 For this case, the needed value for non-slip rule of ${s_3}  $ is 1.6. As shown in Fig.~\ref{FIG1}, the relative error for each method changes with different values of ${s_3}  $. But only in the present method, the minimum value of $ {E_2} $ is achieved when ${s_3}=1.6  $. And when the non-slip condition is satisfied, the relative error reaches a quite small value even in a very coarse mesh.
 
 To further confirm the consistent non-slip boundary condition in the present method, we conduct several others cases. Now the channel width is set to be 50 nodes, and different body forces $ {F_x} = [1 \times {10^{ - 6}},3 \times {10^{ - 6}},5 \times {10^{ - 6}},7 \times {10^{ - 6}}]
  $ are considered. The configurations are the same as those in \cite{premnath2009incorporating}, ${s_3}  $ is chosen according to the non-slip rule, while other relaxation parameters are 1.754. As shown in Table~\ref{TAB1}, the relative errors for $ M_1 $ are $ O({10^{ - 4}}) $, which is consistent with the results in \cite{premnath2009incorporating}, where the non-slip rule was not considered. Compared with these three methods, the relative errors for the present method are much smaller with 5-6 orders, which confirms the availibility of the non-slip rule in the present method.
  For the differeces between the three methods in this case, it is easy to analyze that the error terms in 
  the three methods are in a descending order of $ M_1 $, $ M_2 $, and $ M_3 $. 
   
\begin{table}
	\renewcommand\arraystretch{1.2}
	\caption{\label{TAB1}%
		$ {E_2}( \times {10^4}) $ with different magnitudes of force
		for different methods.
	}
	\begin{ruledtabular}
		\begin{tabular}{lcccr}
			\textrm{$ {F_x}
				 $}&
			\textrm{$ M_1 $}&
			\textrm{$ M_2 $}&
			\textrm{$ M_3 $}&
			\textrm{$ M_p $}\\
			\colrule
		$ 1 \times {10^{ - 6}} $ & 2.739 & 2.339 & 1.113&$ 1.044 \times {10^{ - 6}}
		 $\\
		 
		$ 3 \times {10^{ - 6}} $  & 2.739 & 2.339 & 1.113&$ 3.173 \times {10^{ - 6}}$\\
		
		$ 5 \times {10^{ - 6}} $ & 2.739 & 2.339 & 1.113&$ 5.527 \times {10^{ - 6}}$\\
		$ 7 \times {10^{ - 6}} $ & 2.739 & 2.339 & 1.113&$ 7.296 \times {10^{ - 6}}$
		\end{tabular}
	\end{ruledtabular}
\end{table}

\subsection{Steady Taylor-Green flow}\label{sec.4b}
 
 For the two-dimensional steady incompressible flow in a periodic box $ N \times N
  $, if the force field  is given by,
  \begin{equation}
{\bf{F}}(x,y) = 2\nu {u_0}{\phi ^2}\left[ {\cos (\phi x)\cos (\phi y),\sin (\phi x)\sin (\phi y)} \right],
 \end{equation}
the  flow has the following analytical solution,
\begin{equation}
\begin{array}{l}
{\bf{u}}_a(x,y) = {u_0}[\sin (\phi x)\sin (\phi y),\cos (\phi x)\cos (\phi y)],\\
p_a(x,y) = {p_0} + 0.25u_0^2[\cos (2\phi x) - \cos (2\phi y)],
\end{array}		
\end{equation}
where $ \phi  = 2\pi /N $, $ {p_0} = {\rho _0}c_s^2 $ and $ {\rho _0} = 1 $. The flow is known as steady Taylor-Green flow or four-rolls mill \cite{taylor1934formation}, and is characterized by Reynolds number, $ {\mathop{\rm Re}\nolimits}  = {u_0}\pi /v $.
\begin{figure}
	\includegraphics[width=0.4\textwidth]{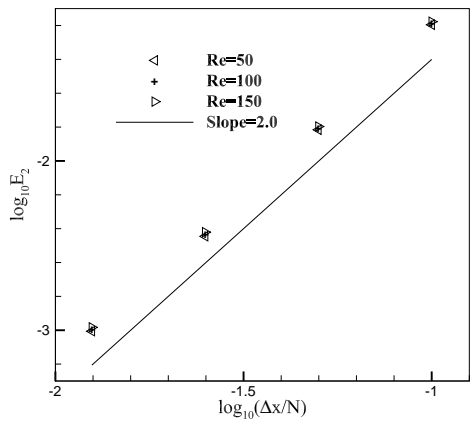}
	\caption{\label{FIG2} $ {E_2} $ changes with grid size  for the present scheme at $ {\mathop{\rm Re}\nolimits}  = 50,100$ and $150
		 $. }
\end{figure}
\begin{table*}
	\renewcommand\arraystretch{1.3}
	\caption{\label{TAB2} $ {E_2}( \times {10^2}) $  and convergence rates (CR) for different forcing methods with different values of Reynolds number.
	}
	\begin{ruledtabular}
		\begin{tabular}{ccccccccccccc}
			&\multicolumn{4}{c}{${\mathop{\rm Re}\nolimits}  = 50$}  &\multicolumn{4}{c}{${\mathop{\rm Re}\nolimits}  = 100$} 
			&\multicolumn{4}{c}{${\mathop{\rm Re}\nolimits}  = 150$} 
			\\ 
			\cline{2-5} \cline{6-9} \cline{10-13}
			
			$ N/\Delta x
			 $&$ M_P $&$ M_1 $&$ {M_2} $&$ M_3 $
			&$ M_P $&$ M_1 $&$ {M_2} $&$ M_3 $
			&$ M_P $&$ M_1 $&$ {M_2} $&$ M_3 $
			\\ 
			\hline			
		10&6.3752&6.5748&6.5522&6.3959&6.5448&6.6456&6.6398&6.5660&6.6482&6.7162&6.7135&6.6698\\
		20&1.5275&1.6263&1.6051&1.5538&1.5788&1.6283&1.6227&1.6063&1.5974&1.6305&1.6279&1.6261\\
		40&0.3587&0.3969&0.3782&0.3843&0.3719&0.3961&0.3909&0.4074&0.3796&0.3957&0.3933&0.4145\\
		80&0.0986&0.1141&0.1037&0.1498&0.1025&0.1109&0.1076&0.1543&0.1040&0.1098&0.1082&0.1560\\
	  CR &2.0133&1.9579&2.0031&1.8264&2.0076&1.9753&1.9896&1.8213&2.0068&1.9848&1.9917&1.8222\\					
		\end{tabular}
	\end{ruledtabular}
\end{table*}
In the simulation, the computational domain covered by a series of grid nodes, $ N/\Delta x = \left[ {10,20,40,80} \right]$, with three different conditions at $ {\mathop{\rm Re}\nolimits}  = [50,100,150] $. To weaken the artificial compressibility, $ {u_0}=0.05 $ is used in all the cases, $ {s_b} $ is given equal to $ {s_2} $, while the remaining relaxation parameters are set to unity. The relative error $ {E_2}
$ is computed from Eq. (\ref{e24}). The relationship between grid size
and  $ {E_2} $ of the present forcing scheme at different Reynolds numbers is presented in Fig.~\ref{FIG2}. The slops at $ {\mathop{\rm Re}\nolimits}  = 50,100$ and $150 $ are 2.0133, 2.0076 and
2.0068, respectively. This demonstrates the scheme preposed has second-order accuracy in space. The relative error  for each method is shown in Table~\ref{TAB2}. It is found that the present scheme achives 
the smallest relative error for every grid resolution at every Reynolds number. Due to the discrete equilibriun central moments used in $ M_3 $ (see Eq. (10) in \cite{de2017non}), some additional errors are introduced into the CLBM, and this effect becomes evident when the mesh size is small. It is the reason why this method manifests an outlier for the finest grid resolution. Generally, each method presents  a second-order convergence rate.

\subsection{Single static droplet}\label{sec.4c}
To validate the availability of the present forcing scheme for a complex force field. We consider the simulations of a statice droplet using the 
Shan-Chen multiphase model \cite{shan1993lattice}, which is also known as
the pseudopotential approach in the multiphase flow. The interaction force is calculated from an interaction potential $ \psi (x) $ \cite{shan1993lattice},
\begin{equation}
 {\bf{F}} =  - G\psi (x)\sum\nolimits_i {{w_i}} \psi (x + {e_i}\Delta t){e_i}
\end{equation}
where $ G $ is used to control the interaction strength and $ {w_i} $ 
are the weights. When only the nearest-neighbor interactions are considered on the D2Q9 lattice, $ {w_i}=1/3 $ for $ {\left| {{e_i}} \right|^2} = 1
 $ and $ {w_i}=1/12 $ for $ {\left| {{e_i}} \right|^2} = 2 $. The exponential form of the pseudopotential is used, i.e., $ \psi (\rho ) = {\psi _0}\exp ( - {\rho _0}/\rho ) $. 
 \begin{figure}
 	\includegraphics[width=0.16\textwidth]{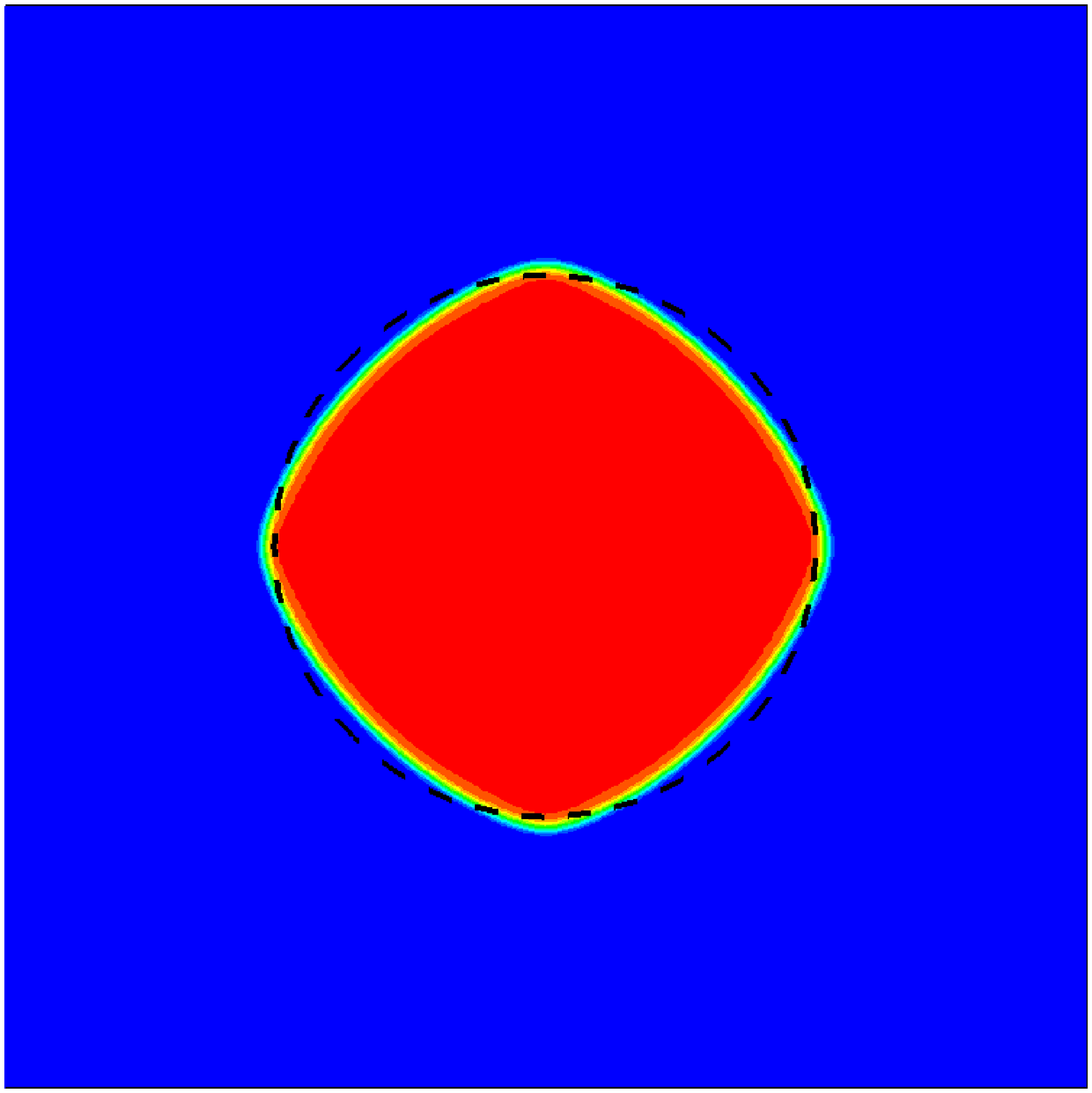}
 	\includegraphics[width=0.16\textwidth]{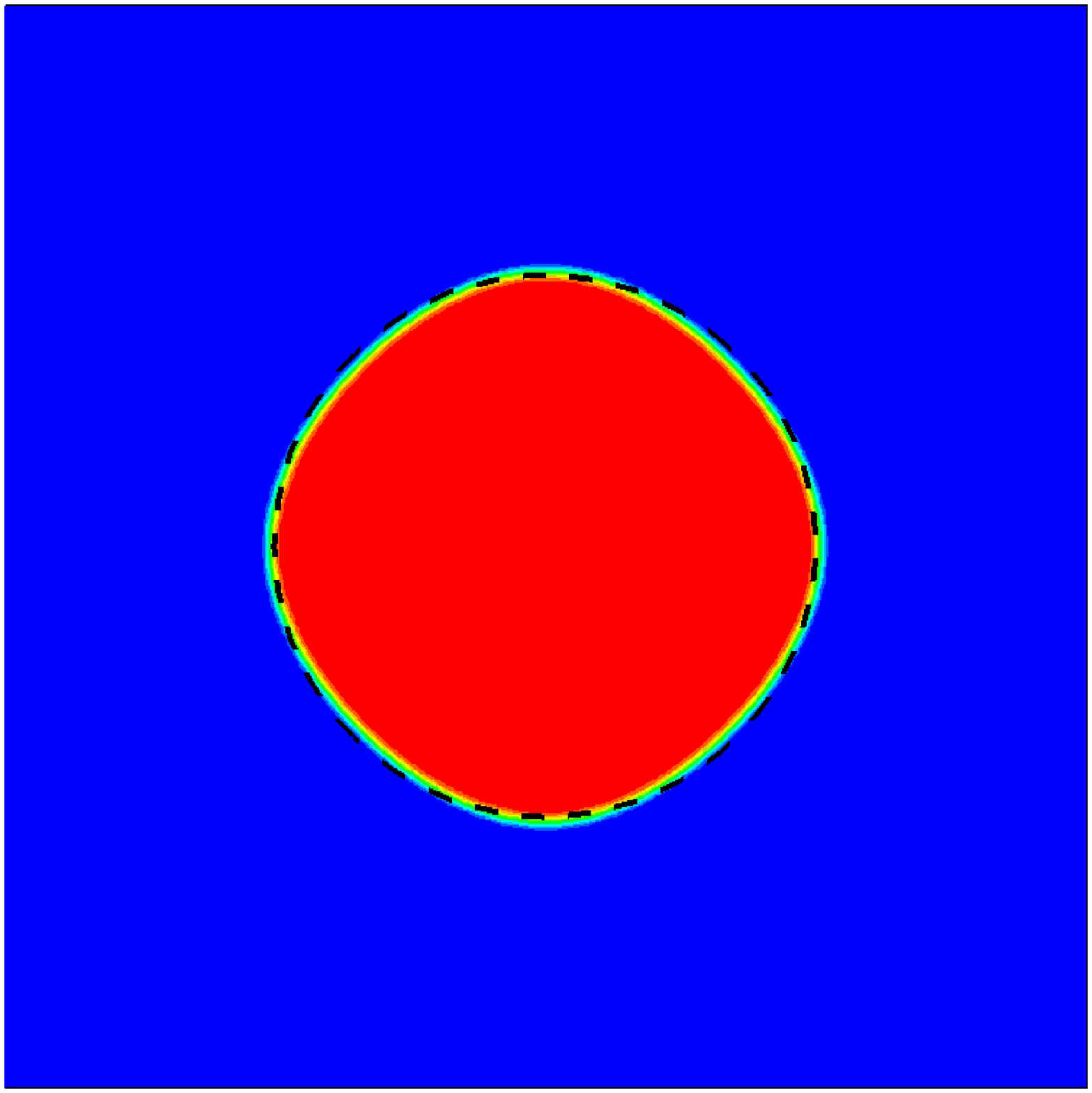}%
 	\includegraphics[width=0.16\textwidth]{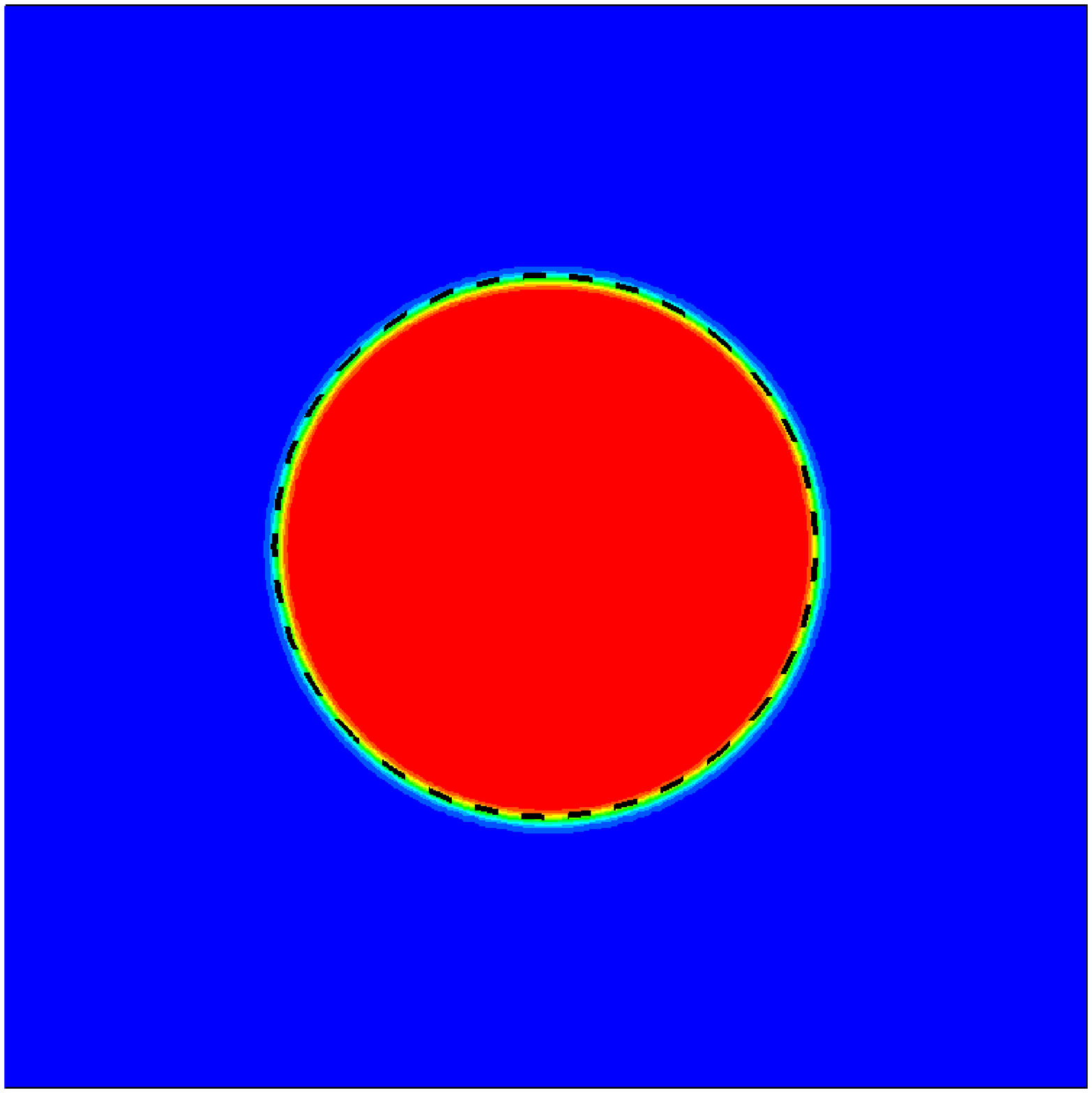}%
 	\caption{\label{FIG3} Steady-state density contours given by $ M_1 $ with
 	$ {s_3} =  [1.0,1.4,1.8] $. }
 \end{figure}
Let us denote $ {\rho _V}
$ and $ {\rho _L} 
$ as the vapor and liquid coexistence densities, respectively.
In this study, $ {\psi _0} = 1 $, $ {\rho _0} = 1$ and $ G = 10/3
$ are used, which leads to $ {\rho _V}=0.3675
$ and $ {\rho _L}=2.783$ \cite{yu2009interaction,li2016revised}. The simulations are conducted in a periodic box $ N \times N =200\times200$. A circle droplet of radius $ R $ is initialized by setting $ \rho  = {\rho _L} $ in the circle and $ \rho  = {\rho _V} $ outside the circle.
The relaxation parameters are chosen as $ {s_b} = {s_v} = 1.4
$, and $ {s_3} = [1.0,1.4,1.8] $.
 \begin{figure}
	\includegraphics[width=0.16\textwidth]{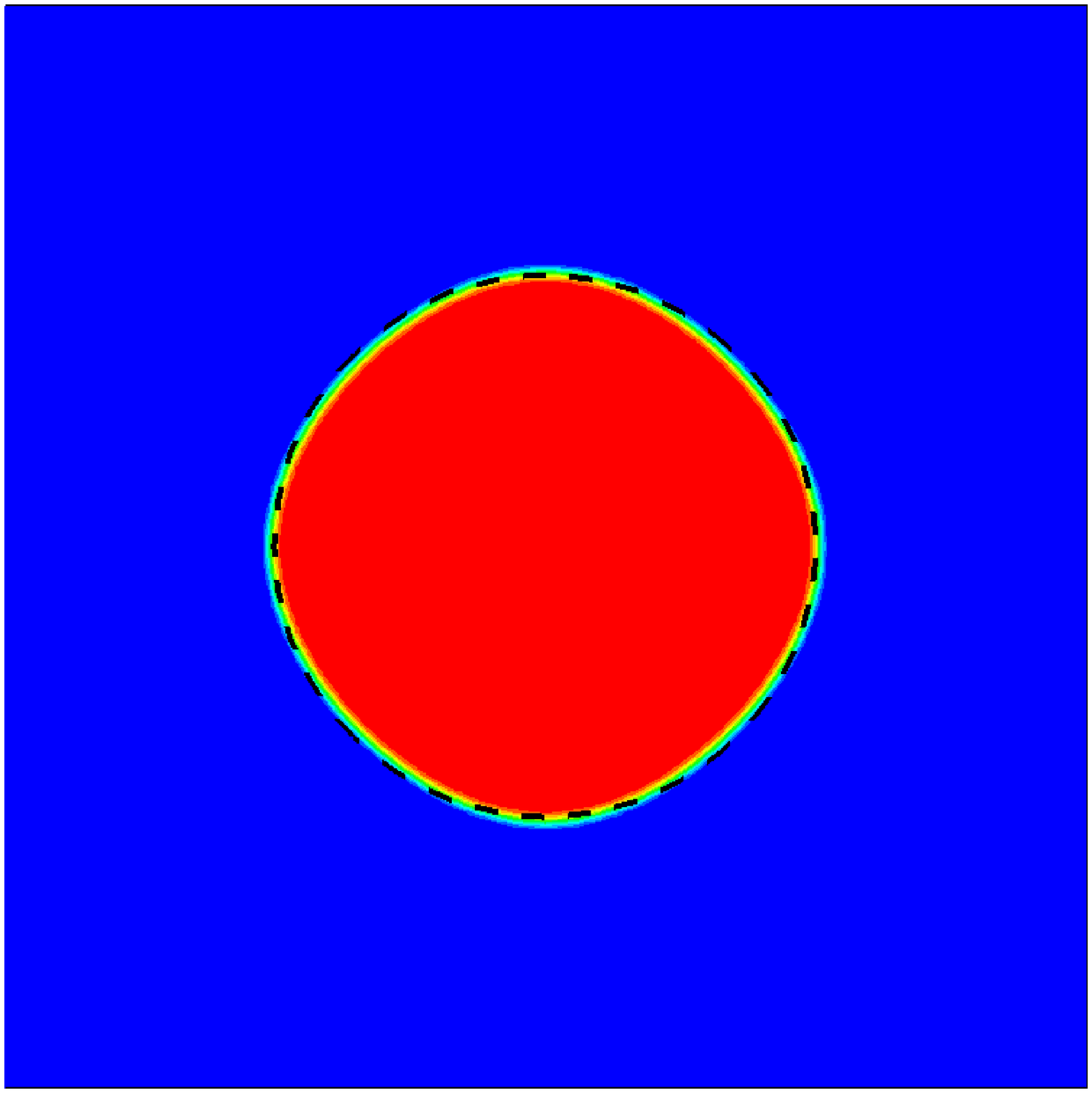}
	\includegraphics[width=0.16\textwidth]{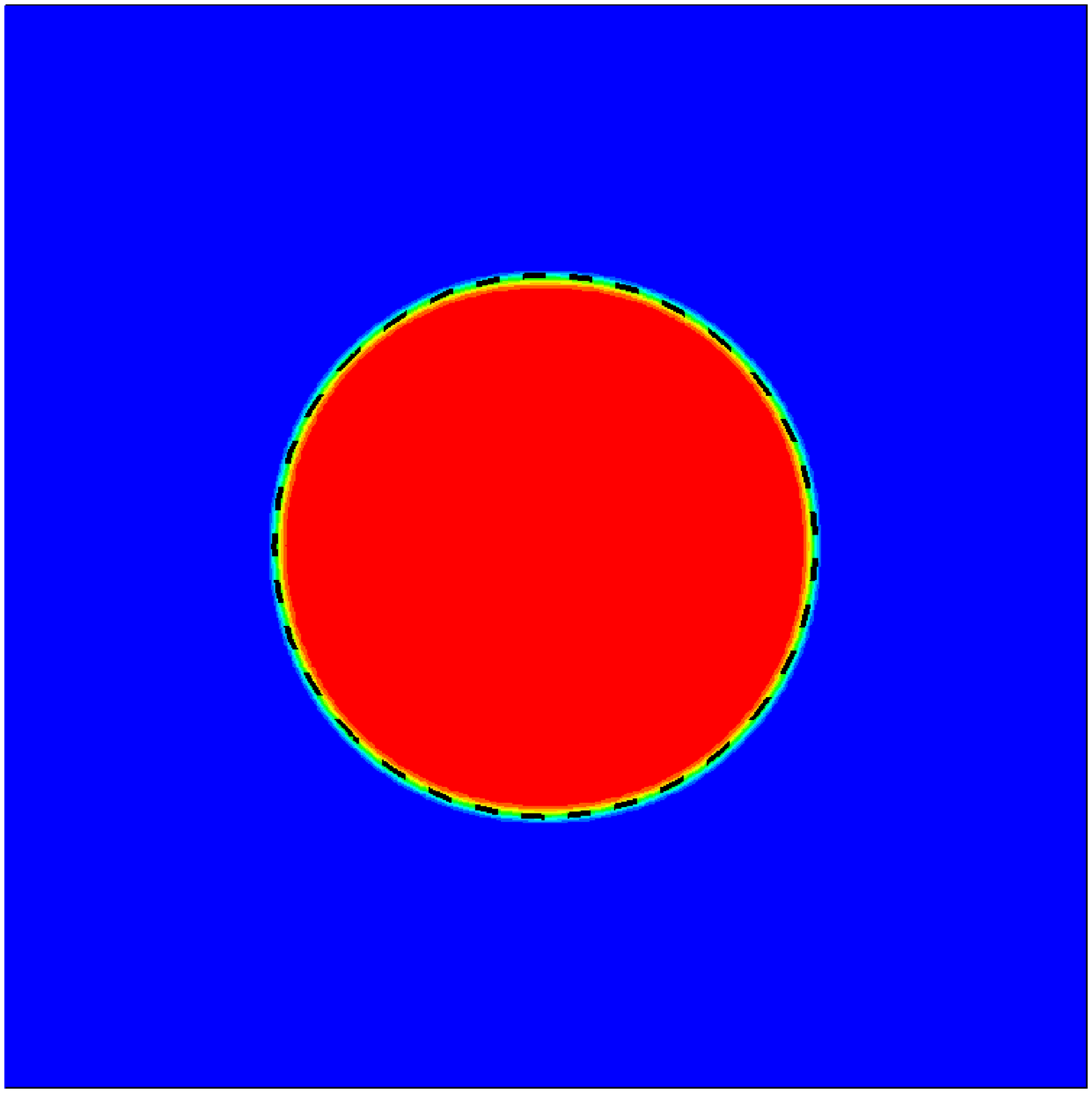}%
	\includegraphics[width=0.16\textwidth]{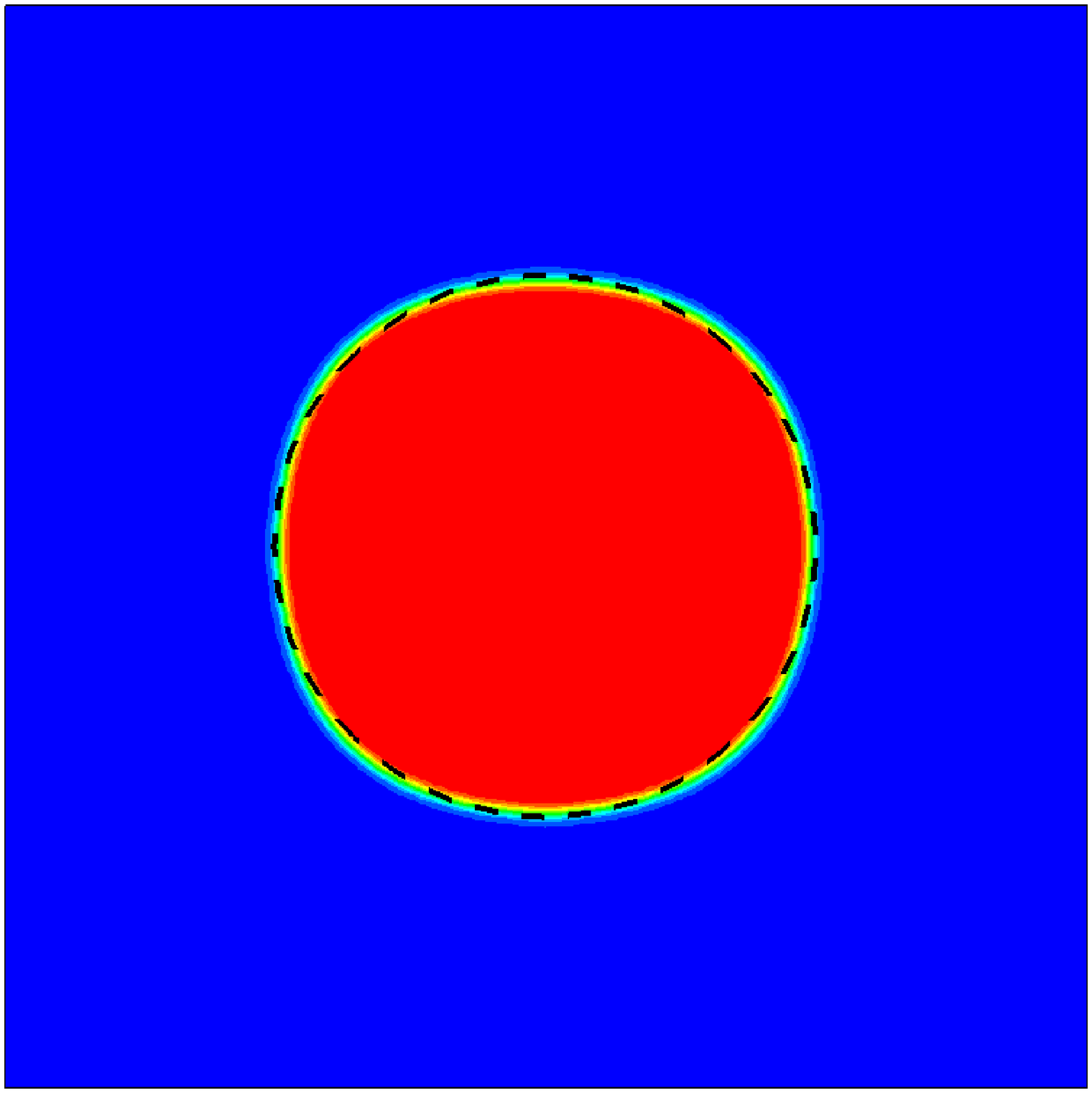}%
	\caption{\label{FIG4} Steady-state density contours given by $ M_2 $ with
		$ {s_3} =  [1.0,1.4,1.8] $. }
\end{figure}
 \begin{figure}
	\includegraphics[width=0.16\textwidth]{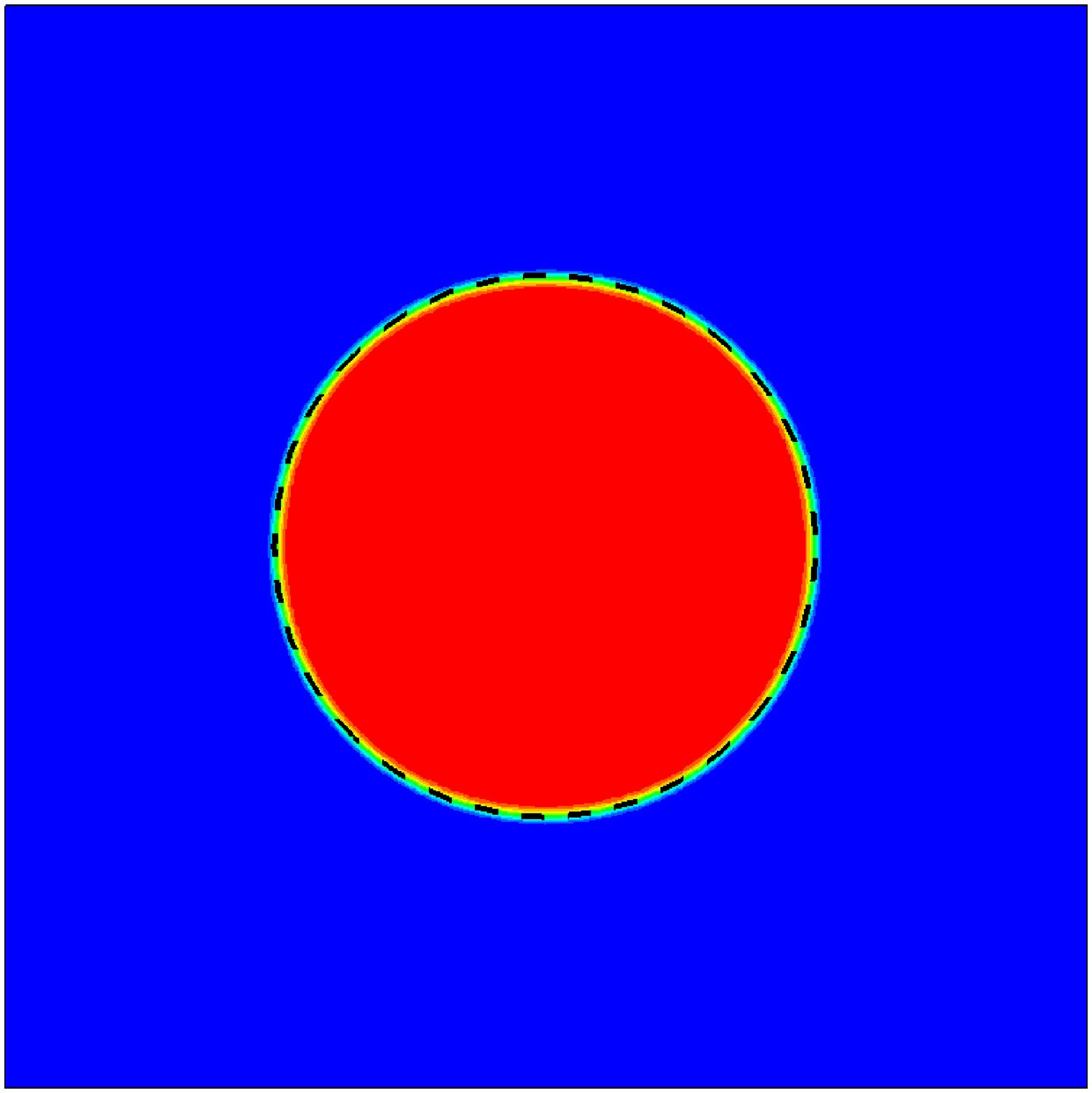}
	\includegraphics[width=0.16\textwidth]{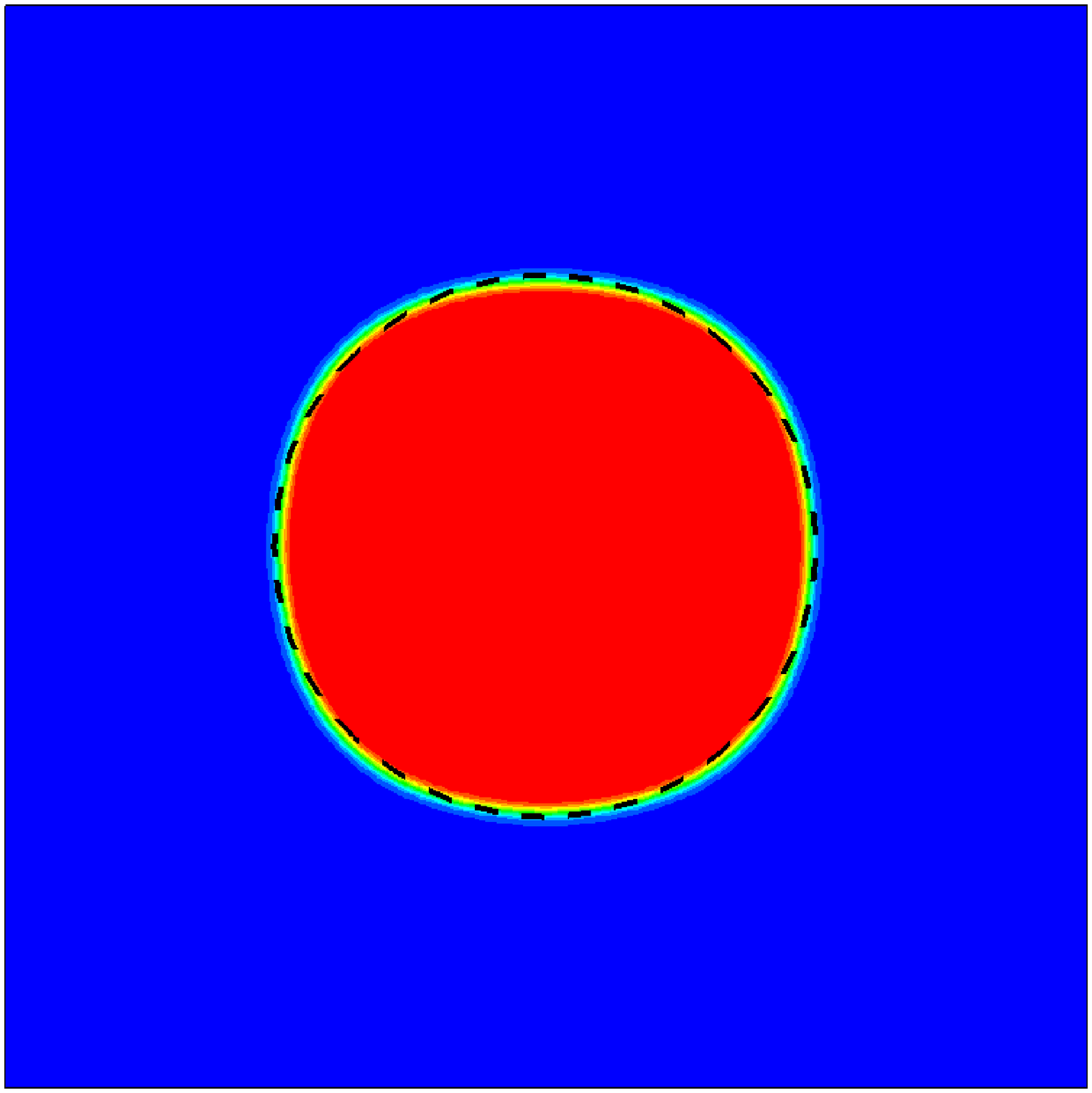}%
	\includegraphics[width=0.16\textwidth]{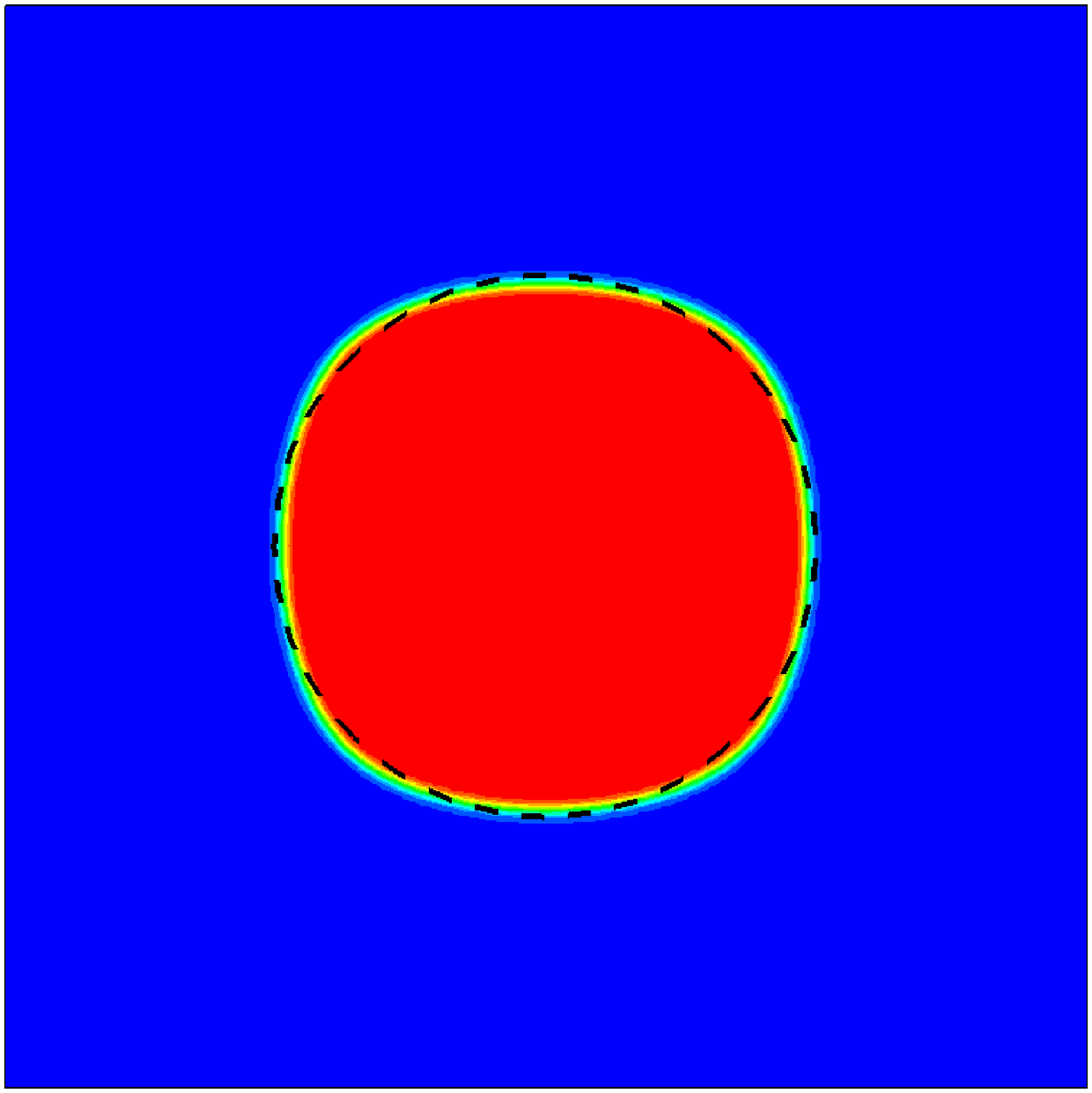}%
	\caption{\label{FIG5} Steady-state density contours given by $ M_3 $ with
		$ {s_3} =  [1.0,1.4,1.8] $. }
\end{figure}

\begin{figure}
	\includegraphics[width=0.16\textwidth]{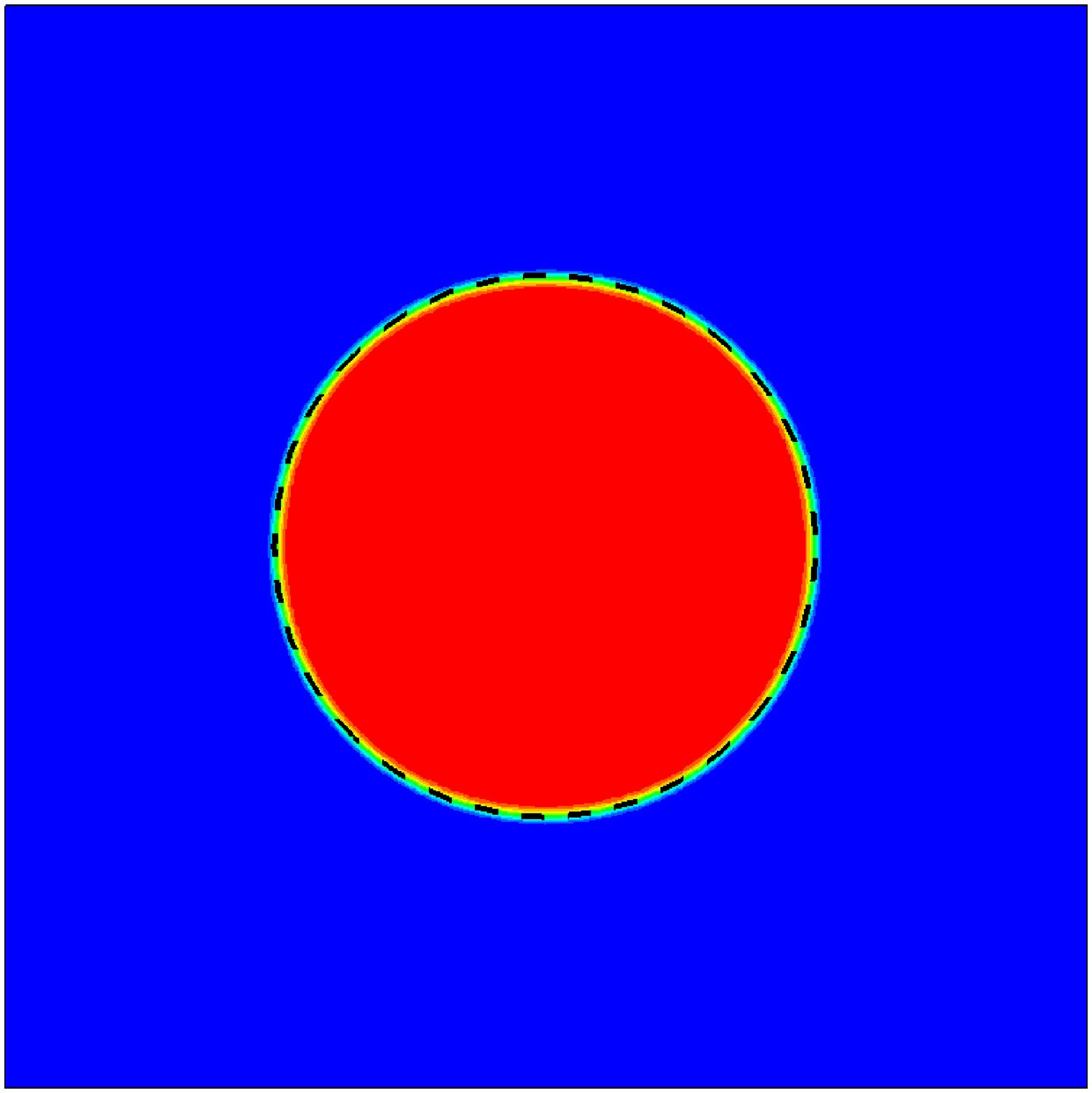}
	\includegraphics[width=0.16\textwidth]{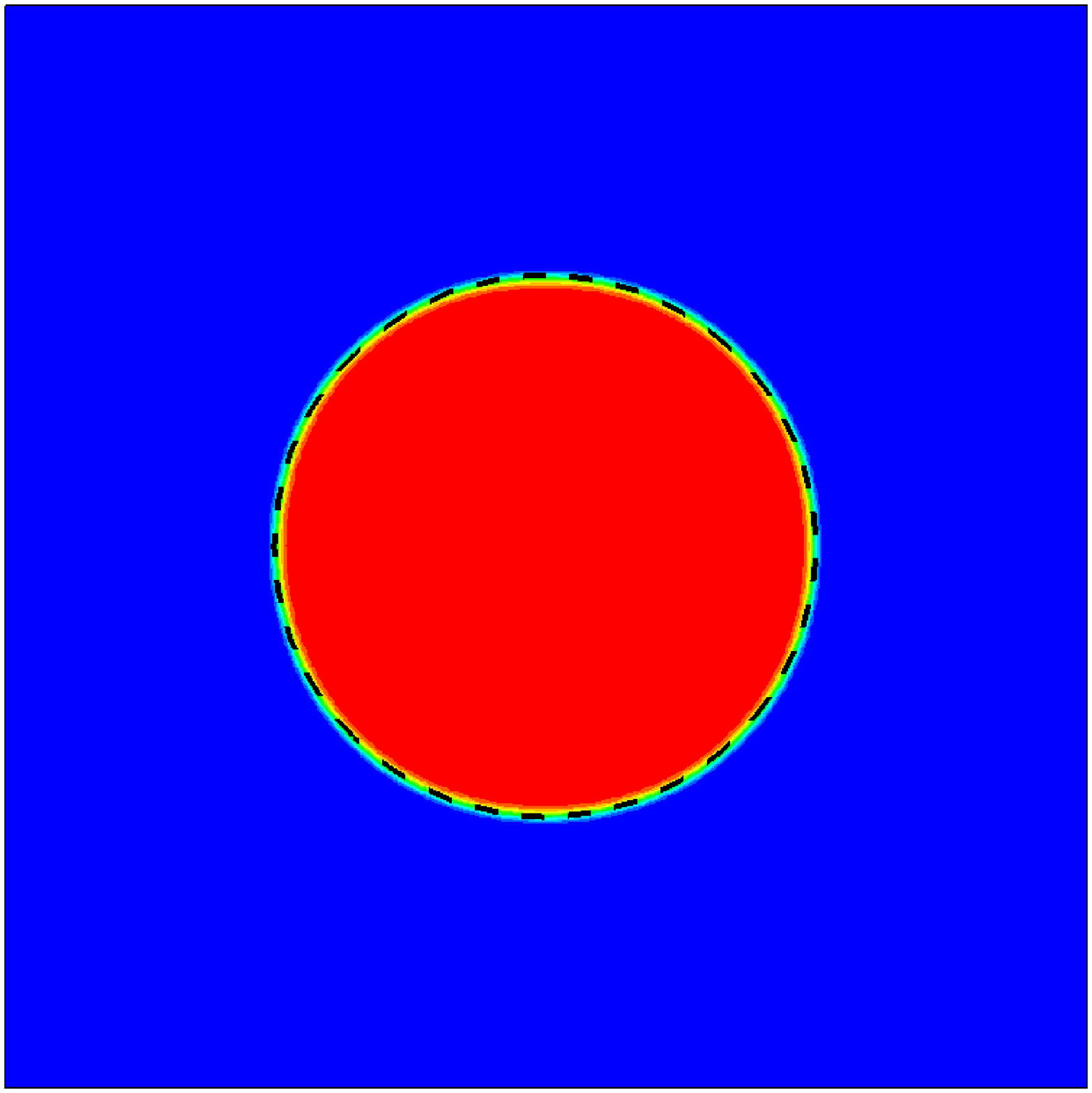}%
	\includegraphics[width=0.16\textwidth]{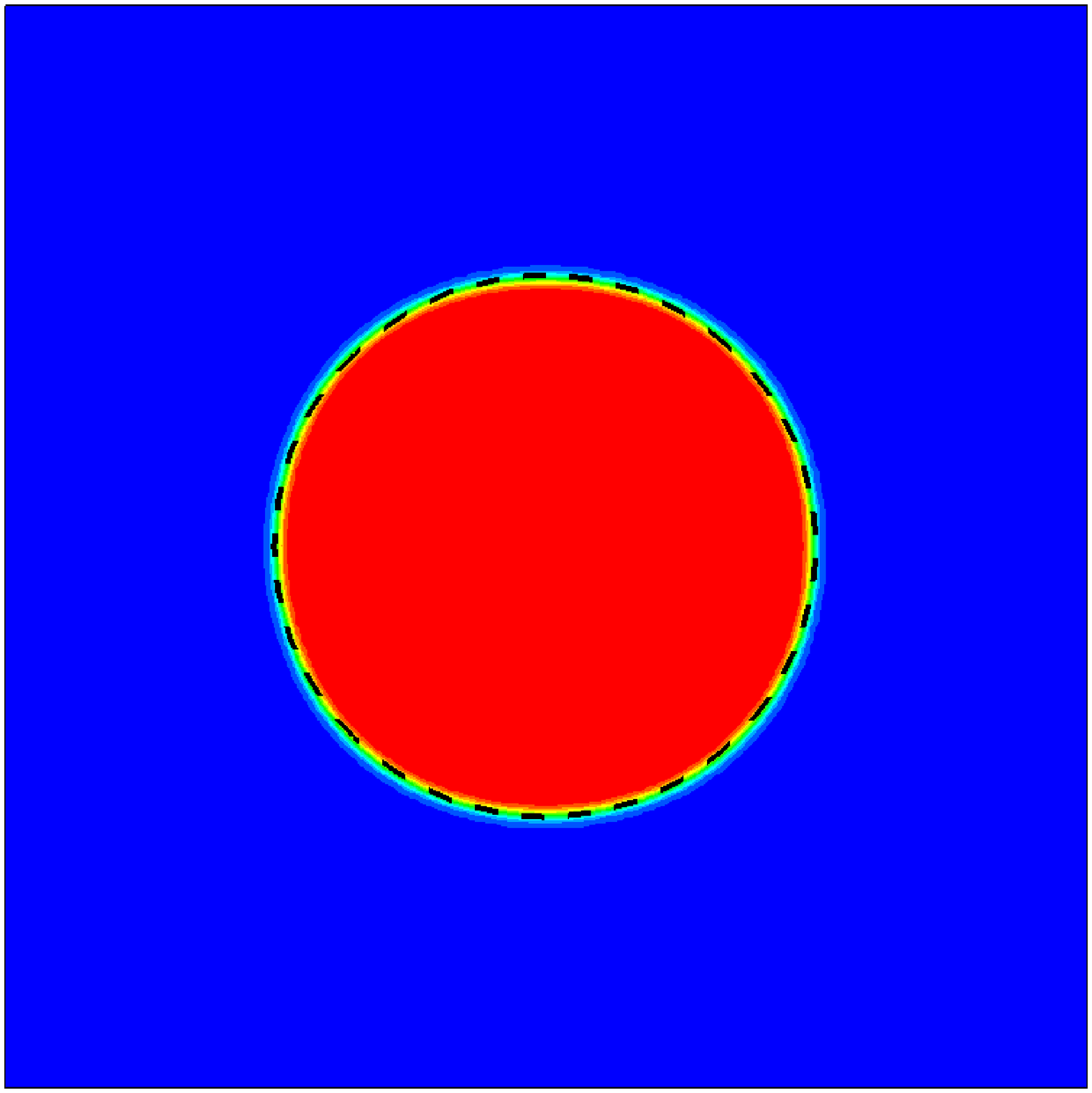}%
	\caption{\label{FIG6} Steady-state density contours given by present method with
		$ {s_3} =  [1.0,1.4,1.8] $. }
\end{figure}

Firstly, the steady-state density contours with $ R=50 $ given by different forcing schemes are compared. The additional dashed circle represents the theoretical location of the droplet. It is found in Fig.~\ref{FIG3} the shape of droplet is ${s_3}$-dependent and it changes from a out-of-round shape to a circle with the increase of $ {s_3} $. As discussed in Sec. \ref{sec.3a}, the removement of high-order terms for the central moments of $ \Delta f
$ in \cite{premnath2009incorporating} makes inconsistencies with the scheme proposed by Guo 
\emph{et al}. And only when $ {s_3} $ is set to be 2.0, the inconsistency can be eliminated.
Though we can not give the result with $ {s_3}=2.0 $ (divergent for this simulation), the tendency confirms our argument. Anologously, as discussed in  Sec. \ref{sec.3b} and  Sec. \ref{sec.3c}, the inconsistencies in $ M_2 $ and $ M_3 $ can only be eliminated under the conditions of  $ {s_3}={s_2} $ and $ {s_3}=1.0 $, respectively. Thus the droplets  in Fig.~\ref{FIG4} and Fig.~\ref{FIG5} become out-of-round when  $ {s_3} $ is not set the specific value. For the present forcing scheme, the droplets are always in round shapes rather than depend on the value of $ {s_3} $, as seen in Fig.~\ref{FIG6}.
 \begin{figure}
 	\includegraphics[width=0.36\textwidth]{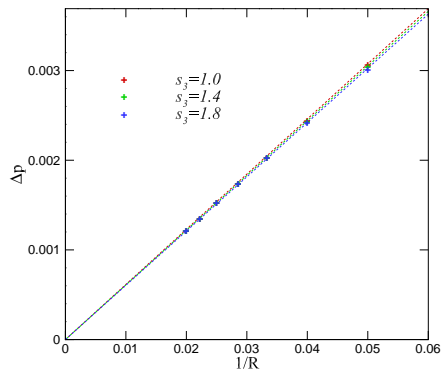}
 	\caption{\label{FIG7} Numerical validation of Laplace's law for the present forcing scheme. }
 \end{figure}
According to Laplace's
law, the pressure difference between the pressure inside and the one outside a droplet is related to the surface tension $ \gamma  $ and the droplet radius $ R $ via $ \Delta p = \gamma /R$.
To check the ability of repeating the Laplace's law, a series of static droplets with $ R = \left[ {20,25,30,35,40,45,50} \right]
$ are simulated. The pressure is computed through $ p = \rho c_s^2 + G{\psi ^2}/2 $. As shown in Fig.~\ref{FIG7}, Laplace's law is well satisfied. The measured
surface tensions for $ {s_3} =  [1.0,1.4,1.8] $ are 0.0615, 0.0610 and 0.0605, respectively.

\section{CONCLUSIONS}\label{sec.5}
In this study, we present a more pellucid derivation of CLBM. A shift matrix $ {\bf{N}}$ is defined in the derivation, by which the raw moments of the discrete distribution function are shifted to their central moments. This definition clarifies the relationship between the MRT LBM and CLBM. Based on this, a new method of incorporating forcing terms into the CLBM is proposed. 

The forcing effect term is incorporated by means of central moments, which is compatible with the basic ideology of the CLBM. According to the definition of the shift matrix $ {\bf{N}}$, the CLBM degrades into the MRT LBM when $ {\bf{N}}$ is a unit matrix. The present forcing scheme retains the property of and  degrades into the Guo forcing scheme in the MRT LBM when  $ {\bf{N}}$ is a unit matrix. Specifically, the present forcing scheme degrades to the original forcing schene proposed by Guo \textit{et al} when all the relaxation parameters are set to the same. Numerical simulations for several benchmark problems confirm the applicability of the non-slip rule, the second-order accuray in space and the property of isotropy for the present scheme. In the meantime, some inconsistences in the previous models are also revealed.

The method developed is quite pellucid, and no cumbersome operations are involved in the practical implementation. Further work will demonstrate that the present scheme can be extended to three dimentions (3D) readily.

\begin{acknowledgments}
Support from the MOST National Key Research and Development Programme (Project No. 2016YFB0600805) and the Center for Combustion Energy at Tsinghua University is gratefully acknowledged. Supercomputing time on ARCHER is provided by the “UK Consortium on Mesoscale Engineering Sciences (UKCOMES)” under the UK Engineering and Physical Sciences Research Council Grant No. EP/L00030X/1.
\end{acknowledgments}

\appendix
\section{Appendixes}
Analogously, the raw moments can be transfromed to the discrete DF through $ {\bf{M}}^{ - 1} $, and the central moments can be shifted
to raw moments through $ {\bf{N}}^{ - 1} $,
\begin{equation}
\begin{array}{l}
\left| {{f_i}} \right\rangle  = {{\bf{M}}^{{\rm{ - }}1}}\left| {{T_i}} \right\rangle,\\ 
\left| {{T_i}} \right\rangle  = {{\bf{N}}^{{\rm{ - }}1}}\left| {{{\tilde T}_i}} \right\rangle .\\ 
\end{array}
\end{equation}
The explicit expressions for $ {\bf{M}}^{ - 1} $ and $ {\bf{N}}^{ - 1} $ are
\begin{widetext}
\begin{equation}
{\bf{M}}^{ - 1} = \left[ 
\begin{array}{c c c c c c c c c}
1 &0 &0&-1&0&0&0&0&1\\
0&1/2&0&1/4&1/4&0&0&-1/2&-1/2\\
0&0&1/2&1/4&-1/4&0&-1/2&0&-1/2\\
0&-1/2&0&1/4&1/4&0&0&1/2&-1/2\\
0&0&-1/2&1/4&-1/4&0&1/2&0&-1/2\\
0&0&0&0&0&1/4&1/4&1/4&1/4\\
0&0&0&0&0&-1/4&1/4&-1/4&1/4\\
0&0&0&0&0&1/4&-1/4&-1/4&1/4\\
0&0&0&0&0&-1/4&-1/4&1/4&1/4\\
\end{array} 
\right],
\end{equation}
\end{widetext}
and 
\begin{widetext}
	\begin{equation}
	{\bf{N}}^{-1} = \left[ 
	\begin{array}{c c c c c c c c c}
	1 &0 &0&0&0&0&0&0&0\\
	{u_x}&1&0&0&0&0&0&0&0\\
	{u_y}&0&1&0&0&0&0&0&0\\
	u_x^2+u_y^2&2{u_x}&2{u_y}&1&0&0&0&0&0\\
	u_x^2-u_y^2&2{u_x}&-2{u_y}&0&1&0&0&0&0\\
	{u_x}{u_y}&u_y&u_x&0&0&1&0&0&0\\
	u_x^2{u_y}&2{u_x}{u_y}&u_x^2& {u_y}/2
	&{u_y}/2&2u_x&1&0&0\\
	u_y^2{u_x}&{u_y}^2&2{u_x}{u_y}&{u_x}/2	
	&-{u_x}/2&2u_y&0&1&0\\
	u_x^2u_y^2&2{u_x}u_y^2&2{u_y}u_x^2&u_x^2/2+u_y^2/2&u_y^2/2-u_x^2/2&4{u_x}{u_y}&2{u_y}&2{u_x}&1\\
	\end{array} 
	\right].
	\end{equation}
\end{widetext}
\nocite{*}

\bibliography{CFCLBM}

\end{document}